\documentclass[12pt,preprint]{aastex}
\def\lax {\ifmmode{_<\atop^{\sim}}\else{${_<\atop^{\sim}}$}\fi}  
\def\gax {\ifmmode{_>\atop^{\sim}}\else{${_>\atop^{\sim}}$}\fi}  
\def\gtorder{\mathrel{\raise.3ex\hbox{$>$}\mkern-14mu
             \lower0.6ex\hbox{$\sim$}}}
\begin{document}

\title{ The Observed Galactic Annihilation Line. Possible Signature of 
the Cluster for  Accreting Small Mass Black Holes}

\author{Lev Titarchuk\altaffilmark{1,3} and  Pascal Chardonnet\altaffilmark{2}}

\altaffiltext{1}{George Mason University/Center for Earth
Observing and Space Research, Fairfax, VA 22030; and US Naval Research
Laboratory, Code 7655, Washington, DC 20375-5352; ltitarchuk@ssd5.nrl.navy.mil }

\altaffiltext{2}{Universit\'e de Savoie/ LAPTH /LAPP, 9 Chemin de;
Bellevue, BP110
74941 Annecy-le-Vieux Cedex FRANCE; chardonnet@lapp.in2p3.fr}

\altaffiltext{3}{NASA/ Goddard Space Flight Center/USRA, code 660, Greenbelt  
MD 20771; lev@milkyway.gsfc.nasa.gov}

\begin{abstract}
Compton Gamma Ray Observatory, OSSE, SMM, TGRS, balloon and recent INTEGRAL data reveal  a
feature of the 0.511 MeV  annihilation radiation of the Galactic
Center with a flux of approximately $5\times 10^{-4}~0.511$ MeV photons
cm$^{-2}$s$^{-1}$. 
We argue that $e^{+}e^{-}$ pairs can be generated when the X-ray radiation photons and 
$\sim10-30$ MeV photons interact with each other in the compact region in the proximity  of the Galactic 
Center black hole.
In fact, disks formed near black holes of $10^{17}$ g mass should emit  
the $\sim 10$ MeV temperature blackbody radiation. If  positron ($e^{+}$) sources
are producing about $10^{42}~e^{+}$s$^{-1}$ near the Galactic Center 
they would  annihilate on the way out and result in $0.511$ MeV emission. We suggest that 
the annihilation radiation can be
an observational consequence of the interaction  of the accretion disk 
radiation of the SMall Mass Black Holes (SMMBHs) with X-ray radiation in the Galactic Center. 
This is probably   the only way to identify and observe these SMMBHs. 
\end{abstract}

\section{Introduction}

The Galactic Center (GC) of our Galaxy found in 1974 as a strong radio source called Sgr A$^{\ast}$ is the site 
 a supermassive black hole
weighing 2.6 millions solar mass is present (e.g. Melia and Falcke 2001).
Morever the total mass of stars enclosing the inner 20 pc of the Galaxy is $\sim 10^{8}$ M$_{\odot}$, or $\sim0.1$\% of the
Galactic stellar mass (Baganoff et al. 2003). Therefore, the X-ray observations of the Galactic Center can potentially reveal
 accreting black holes and neutron stars. During five years of Chandra observations, Baganoff et al. identified seven X-ray
transients located within 20 pc of Sgr A$^{\ast}$.  Remarkably four of these transients lie within only 1 pc of Sgr A$^{\ast}$
(Muno et al. 2005). This implies that, compared to the numbers of similar systems located between 1 and 20 pc, transients are
over-abundant by a factor of $\sim20$ per unit stellar mass within 1 pc of Sgr A$^{\ast}$. 
 
It is
also now a subject of intense activity due to the possible presence 
of a dark matter cusp inside the Galactic Center. Numerous 
space missions have observed the center of our Galaxy in 
radio, infrared, X and $\gamma$ rays. 
Thanks to these measurements we have a more precise picture of this 
region.  The center of our Galaxy presents the characteristic that 
there is a strong activity both in X and  $\gamma$ rays. 
A recent HESS (High Energy Stereoscopic System)  experiment reports a detection of very high energy gamma rays from the Galactic center
within $1^{\prime}$ of Sgr A$^{\ast}$. The flux above 165 GeV is order of $1.8 \times10^{-7}$ m$^{-2}$ s$^{-1}$ and consistent with a luminosity of $10^{35}$ ergs s$^{-1}$
(Aharonian, et al.  2004).

The most intense line of our Galaxy is coming also from GC: the  511 keV
annihilation line.
Balloon flight observations reported in  1972 showed  first evidence for the
existence of an annihilation line  from the Galactic Center region
(Johnson, Harnden \& Haymes 1972; see also Johnson \& Haymes, 1973).
Since that time the Galactic Center has been the object of intense
observations from space missions and balloon experiments (Leventhal, MacCallum \& Stang 1978; Purcell et al. 1993; Ramaty, Skibo \& Lingenfelter 1994;
Purcell et al. 1997; and Cheng et al. 1997; Share et al. 1990; Teegarden et al. 1996;  Gehrels et al. 1991; Leventhal et al. 1993; 
 Niel et al. 1990;  Chapuis et al. 1991). 

 The observed emission of the  brightest gamma-ray line in our Galaxy is 
due to two separate sources: a diffuse one and a point source near the
Galactic Center. For the point source 
we use the  value given by INTEGRAL from  recent observations (Churazov et al. 2005).
The point source line flux is : $9.9 \times 10^{-4}$~ph~cm$^{-2}$~s$^{-1}$, for which the
corresponding luminosity of the line emission is about 
$6 \times 10^{36}$ ergs~s$^{-1}$ for the distance of about 8 kpc. Then the positron 
sources are produced at a rate given by: 
$\sim  10^{42} d^2_{8}~$s$^{-1}$, where 
$d_{8}$ is the distance expressed in unit of 8 kpc. This
observation is in a good agreement with the previous measurements and is
well fitted by a gaussian with the full width at half maximum of $\sim$ 
10 degrees, with a 2 $\sigma$ confidence interval of 6-18 degrees.\
The  INTEGRAL  measurement providing the line width of  $ \sim 2.4$ keV  exlcudes velocity $v$
of the interstellar medium (ISM) greater than $ \sim 800 $ km s$^{-1}$, i.e. $\beta=v/c \le 2 \times
10^{-3}$. 
 
Several possible sources of the observed line related to the  positron sources  have been proposed. 
For example, Dermer \& Skibo (1997) suggest that  these sources could be associated with galactic fountains of radioactive
elements  ejected from supernova activity in the Galactic Center.
Boehm, Fayet \& Silk (2004) offer an alternative explanation that the observed annihilation line could be
 related to the annihilation of  dark light matter particles into $e^+ e^-$.
In fact, Guessoum, Jean \& Gillard (2005) present the list of the main processes that led  to the positron production.
They are: (i). the $\beta^{+}$ decay of a radioactive nucleous which is widespread in the explosive and/or hydrostatic nucleosynthesis
enviroments of novae, supernovae, Wolf-Rayet, and Asymptotic Giant Branch stars, (ii). $\pi^{+}$ decay into $\mu^{+}$ and gives off a
positron taking place where the pion is produced in collisions of highly energetic ($E>200$ MeV) cosmic rays with interstellar
material, (iii). pair production by interaction of an electron with a strong magnetic field and (iv) the process of the photon-photon pair
production which requires photons of high energies  occuring in the enviroment of luminous compact objects, black holes, active
nuclei (AGN) an etc.
Thus one cannot exclude {\it that pair production is a result of the interaction of X-ray and $\gamma-$ emission
in the very compact shell near Galactic Center}.
 
X-ray observations of the Galactic Center have been made with several instruments (Watson et al. 1981; Skinner et al. 1987;
Kawai et al. 1988; Takeshima et al. 1991; Sunyaev et al. 1993; Markevitch et al. 1993; Pavlinsky et al. 1993; Preddehl, \& Trumper 
1994). High-sensitivity imaging observations with  the Einstein Observatory (Watson et al.) and
ROSAT (Preddehl, \& Trumper) resolved  several discrete sources within $10^{\prime}$ of Galactic Center. One of 
the ROSAT sources, RXJ 1745.6-2900 (Preddehl \& Trumper), is within $10^{\prime\prime}$ coincident of the nonthermal 
radio source Sgr A $^{\ast}$, which is believed to be the dynamical center of our Galaxy.  

The Japanese satellite
GINGA having detected K$\alpha $ lines from highly ionized iron,
found a large amount of hot plasma which is strongly concentrated towards the Galactic Center (Koyama et al. 1989; Yamauchi et al.
1990).
The ASCA image (Tanaka et al. 1995) has revealed extended diffuse emission over the Galactic Center region. 
The observed spectrum shows the K$_{\alpha}$ lines from highly-ionized ions of silicon, sulphur, argon, calcium, in
addition to the high-ionization iron lines. This gives firm evidence for the presence of high-temperature ($kT\approx10$ keV) plasma.
A large energy generation rate of $\sim 10^{41-42}$ ergs s$^{-1}$ is required to produce the observed plasma.
In fact,  Koyama et al. (1996) argue for a source at the Galactic Center.
The luminosity required to account for the 6.6-keV line from  Sgr B2 (the prominent molecular cloud)  is $2\times 10^{39}$ ergs s$^{-1}$
(in 2-10 keV band). While the Sgr A region has  an X-ray luminosity of $\sim 10^{36}$ ergs s$^{-1}$ at present, it could have been much
more luminous $\sim 300$ yr ago. The distinctly bright region inside the Sgr A shell is probably due to this recent activity 
and  the X-ray luminosity may have been  as high as   several  $\times 10^{39}$  ergs s$^{-1}$.  Koyama et al. conclude that  the ASCA
results suggest the presence of an active galactic nucleus  and together form  the evidence for a large mass concentration at the Galactic
Center and transient activities due to sporadic mass accretion rate. 
Muno et al. (2004a) also presented the study of the spectrum of diffuse emission in $17^{\prime}\times 17^{\prime}$ field around 
Sgr A$^{\ast}$ during 625 ks of Chandra observations. They confirmed and extended the results of the previous study (see e.g. Koyama et al.).
In particular, they showed that the contiuum and line spectra are consistent with an origin in a two-temperature plasma. As the cooler,
$kT\approx 0.8 $ keV plasma differs in the surface brightness, the hotter, $kT\approx 8$ keV component is more spatially uniform.

Recently Revnivtsev et al. (2004) have  published INTEGRAL data on giant molecular cloud Sgr B2's (IGR J17475-2822) fluorescence. 
They have constructed the broad band (3-200 keV)
spectrum from the ASCA-GRANAT-INTEGRAL data. It  appears that the X-ray irradiating spectrum of the  cloud is, in fact, not free-free, (as  early proposed irradiating spectrum,
see above) and it is  rather quite hard, extending to over 100 keV
and probably a power law with  photon index 1.8.    Revnivtsev et al. concluded that 300-400 years ago Sgr A$^{\ast}$ was a low luminosity (of $\sim 2\times10^{39}$  ergs
s$^{-1}$ at 2-200 keV) AGN with characteristic hard X-ray Comptonization spectrum with photon index of about 1.8 and plasma
temperature is about 50 keV.   


The $\gamma-$  emission 
between 1 MeV and 30 Mev has been measured by COMPTEL. Using the COMPTEL data points (Hunter et al. 1997) one can calculate that 
the luminosity  in 1-30 MeV energy band is about $4\times10^{38}$ ergs s$^{-1}$ with an assumption that the distance to the Galactic Center is about 8 kpc.


In this Paper, we study a possibility of  pair creation due to X-ray and $\gamma-$ photon interaction inside the compact Galactic
Center shell (the SGR A shell) and we discuss a possibility of annihilation line formation when these pairs propagate through the Galaxy from the Galactic Center
region.  
In \S 2 we present a pair creation model. In \S 3 we  compare our theoretical predictions for the pair and
annihilation photon  production with observations. We determine the size of the pair production area using
the observable fluxes in X-ray and $\gamma-$ray energy ranges. In \S 4 we investigate the outward propagation of positrons from the Galactic Center
region.  In \S 5 we present arguments for the origin of
the $\sim 10 $ MeV emission near Galactic Center. We discuss our results in \S 6 and we make our
final conclusions  in \S 7.

\section{Pair creation  model}

We assume that the photon density of X-ray photons emitted by the inner shell near 
the central BH within $10^3$ $R_{\rm S}$ (Schwarzschild radii)  is described by the following  distribution over radius $r$ (see Fig. 1 for the geometrical description of the model)
\begin{equation}
n_X=\frac{{\cal L}_X}{4\pi r^2 E_Xc}
\end{equation}
where ${\cal L}_X$ is  X-ray luminosity of the central source per keV (X-ray energy spectrum), $E_X$ is the energy of
X-ray photon and $c$ is the speed of light.
Also we consider that the spherical shell between the inner radius $R_c$ and 
$R$ ($R_c\leq r\leq R$ and $R\gg R_c$) is uniformly filled by $\gamma-$radiation sources where photon density
production $P_{\gamma}$  is
\begin{equation}
P_{\gamma}=\frac{{\cal L}_{\gamma}}{E_{\gamma}4\pi R^3/3}
\end{equation}
where ${\cal L}_{\gamma}$ is the $\gamma$-radiation luminosity of the spherical shell 
per keV ($\gamma$-ray energy spectrum) and  
$E_{\gamma}$ is the energy of
the $\gamma$-ray photon.
Then the pair production by the photons (for which the pair production condition $E_XE_{\gamma}>(m_ec^2)^2$
 is satisfied) is
\begin{equation}
P_{e^-e^+}^{X,\gamma}=4\pi\int_{R_c}^{R}r^2P_{\gamma}dr \int_{\Omega} T_{X,\gamma}
(r,\varphi,\theta)d\Omega
\end{equation}       
where $T_{X,\gamma}(r,\varphi,\theta)$ is the optical path 
for $\gamma$-ray photon propagating 
in the X-ray background  and incoming at  radius $r$ at  given direction ${\bf \Omega}$. 
The multiple integral in equation (3) can be calculated analytically.
In fact,
$$
\int_{\Omega} T_{X,\gamma}(r,\varphi,\theta)d\Omega=\{2\pi\sigma_{X,\gamma}\}\times
$$
\begin{equation}
\left\{\int_0^1d\mu
\left[\int_{r(1-\mu^2)^{1/2}}^{\infty}n(r^{\prime})ds+
\int_{r(1-\mu^2)^{1/2}}^{r}n(r^{\prime})ds\right]+
\int_{-1}^{0}d\mu \int_{r}^{\infty}n(r^{\prime})ds\right\}
\end{equation} 
where 
\begin{equation}
n(r^{\prime})ds=\frac{n_0r^{\prime}dr^{\prime}}{{r^{\prime}}^2[{r^{\prime}}^2-r^2(1-\mu^2)]^{1/2}}
\end{equation}
and $n_0={\cal L}_X/4\pi E_Xc$; where $\sigma_{X,\gamma}(E_X, E_{\gamma})$ is a pair creation cross section given by (Dirac, 1930,
Heitler, 1954)
\begin{equation}
\sigma_{X,\gamma}(E_X, E_{\gamma})=\frac{\pi {\cal R}_0^2}{2}(1-b^2)\left[2b(b^2-2)+(3-b^4)
\ln\left(\frac{1+b}{1-b}\right)\right]
\end{equation}
and $b=[1-(m_ec^2)^2/E_XE_{\gamma}]^{1/2}$, and where the classical electron radius 
${\cal R}_0=e^2/m_ec^2=2.818\times10^{-13}$ cm.

The analytical integration of
$\int n(r^{\prime})ds$
\begin{equation}
\int n(r^{\prime})ds =n_0\int \frac{dr^{\prime}}{r^{\prime}\sqrt{{r^{\prime}}^2-r^2(1-\mu^2)}}=
\frac{n_0}{r\sqrt{1-\mu^2}}\arccos{\frac{r\sqrt{1-\mu^2}}{r^{\prime}}} +C
\end{equation} 
 allows us to calculate the second integral of $\int n(r^{\prime})ds$  over $\mu$:
\begin{equation}
\int_{\Omega} T_{X,\gamma}(r,\varphi,\theta)d\Omega=
2\pi\sigma_{X,\gamma}\frac{{\cal L}_X}{4\pi E_Xc}\frac{3\pi^2}{4r}.
\end{equation}
Finally, the expression for the Green's function $P_{e^-e^+}^{X,\gamma}$
 can be written as 
 \begin{equation}
P_{e^-e^+}^{X,\gamma}= 
\frac{9\pi^2}{16cR}\frac{\sigma_{X,\gamma}{\cal L}_X{\cal L}_{\gamma}}{E_XE_{\gamma}}.
\end{equation}
In order to obtain the total pair production rate $P_{e^-e^+}$ one should integrate 
$P_{e^-e^+}^{X,\gamma}$ over X-ray and $\gamma-$ energies, namely
\begin{equation}
P_{e^-e^+}=\frac{9\pi^2}{16cR}\int_0^\infty dE_X\int_{(m_ec^2)^2/E_X}^\infty
\frac{\sigma_{X,\gamma}{\cal L}_X{\cal L}_{\gamma}}{E_XE_{\gamma}}dE_{\gamma}.
\end{equation}

\section{Results of calculations}

To compute the positron flux one needs the X and $\gamma $ ray 
energy spectra, ${\cal L}_X$ and ${\cal L}_{\gamma}$ respectively. We will use for 
the total X-ray and $\gamma$
luminosities of the inner GC region,  
$L_{X} \sim 2\times 10^{39}$ ergs s$^{-1}$  and   $L_{\gamma} \sim 4 \times 10^{38}$ ergs s$^{-1}$ respectively (see  a review of the observations 
in the Introduction).

We   assume  that the $\gamma-$ radiation  emanates  from a optically thick medium and consequently the emergent spectrum has a  blackbody like shape that color
temperature $T_{\gamma} =10$ MeV (see Fig. 1). Then for a normalization constant $C_{\gamma}$ we have an equation as follows 
\begin{equation}
L_{\gamma} 
= \int_{0}^{\infty}{\cal L_{\gamma}}(E_{\gamma}) dE_{\gamma}= C_{\gamma}\int_{0}^{\infty} \frac{y^3}{\exp(y) -1} dy
\end{equation}
 where
$y=E_{\gamma}/ kT_{\gamma} $.  For the X-ray photons, we
assume a power law with an exponential cutoff
\begin{equation}
\displaystyle{L_{X} 
= \int_{0}^{\infty} {\cal L_{X}}(E_{X}) dE_{X}= C_{X}\int_{0}^{\infty} x^{-\alpha} \exp(-x)  dx}
\end{equation}
where
$x=E_{X}/ kT_{X}$, $kT_{X}=100$ keV, the spectral index $\alpha=\Gamma-1=0.8$ and $C_{X}$ is the normalization constant. 
From the previous section we saw that
the process of pair creation is driven by the cross section given by
equation (6). The energy dependence, represented by the parameter
$b$  can be expressed using the dimesionless energies $x=E_{X}/ kT_{X}$
and $y=E_{\gamma}/ kT_{\gamma}$, namely $b=[1-(m_ec^2)^2/E_XE_{\gamma}]^{1/2} =[1-3.26/xy]^{1/2}$.
Then the formula for the pair flux 
\begin{equation}
P_{e^-e^+}=\frac{9\pi^2}{16cR}\int_0^\infty dE_X\int_{(m_ec^2)^2/E_X}^\infty
\frac{
\sigma_{X,\gamma}( E_X,E_{\gamma}){\cal L_{\gamma}}(E_{\gamma}) {\cal L_{X}}(E_{X})} 
{E_XE_{\gamma}}dE_{\gamma}
\end{equation}
can be modified  using the variables $x$ and $y$ as follows
\begin{equation}
P_{e^-e^+}=\frac{9\pi^2}{16cR}\int_{0}^{\infty} dx\int_{3.26/x}^\infty
\frac{\sigma_{ E_X,E_{\gamma}}(x,y){\cal L_{\gamma}}(y) {\cal L_{X}}(x)} 
{xy}dy.
\end{equation}
We compute this double integral and obtain the following result:
\begin{equation}
P_{e^-e^+}= \frac{3.2 \times 10^{56}}{R}~~~{\rm cm^{-1}~s^{-1}}.
\end{equation}
To reproduce the value of $10^{42}$s$^{-1}$ for the
positrons rate, one needs a value of the radius $R$,  given by:
\begin{equation}
R=\frac{3.2 \times 10^{56}}{P_{e^-e^+}} 
=\frac{3.2 \times 10^{56}}{10^{42}} =  3.2\times10^{14}~~~{\rm cm}.
\end{equation}

It is worth noting that the intrinsic pair flux $\sim10^{36}$ s$^{-1}$, formed as a result of $\gamma+\gamma$ interaction (to calculate this flux one should replace 
${\cal L_{X}}$ by ${\cal L_{\gamma}}$ in formula 14) in this $\gamma-$emission region of size $3.2\times10^{14}$ cm, is six orders of
magnitude less than  the pair flux $\sim10^{42}$ s$^{-1}$  required to reproduce   the observed annihilation line strength. 
\section{Outward propagation of positrons from the Galactic Center region}

A positron, $e^{+}$ may collide with an electron $e^{-}$, to produce two gamma ray photons according to the
reaction $e^{+}+e^{-}\to\gamma+\gamma$. Before decaying, the positron 
in the free space interacts with an electron and forms the positronium, a
bound state with  lifetime of order of $10^{-10}$ s. Then particles
annihilate into two or three photons. 
One photon will have  a high energy and ,if the electron is at rest;
the other photon will have an energy of the order  $m_ec^2=511$ keV. In our calculations we assume that 
most of the produced positrons are thermalized in the surrounding thermal plasma and thus we can
consider the annihilation of thermal positrons at a temperature of order $10^7$ K , i.e. 
$\beta=<v>/c\sim 10^{-3}$, where $<v>$ is the thermal velocity of electrons. In fact, the recent INTEGRAL observations of 511 keV annihilation line 
(Churazov et al. 2005) exclude velocity of the interstellar medium greater than $10^{-3}c$. For $\beta\ll1$, 
the cross-section $\sigma$, for  positron annihilation with a free electron at rest is given by Dirac (1930)
\begin{equation}
\sigma\approx\pi r_0^2/\beta.
\end{equation} 
Murphy, Dermer \& Ramaty (1987)  showed that the fraction of $e^{+}$ that annihilites in flight prior to
thermalizing usually amounts to less that $10$ \%, and that these positrons do not contribute to the 511 keV line
emission. After entering the thermal pool, the positron annihilate in a fully ionized thermal plasma through direct
annihilation.
The problem of the positron thermalization and annihilation photon production in the interstellar medium  has been recently studied in details by
Guessoum, Jean, \& Gillard (2005) (see for a review of this subject). 
They reexamine in the utmost detail the various processes undergone by positron annihilation using most recent interaction cross section with atomic and molecular hydrogen,
as well as helium. Putting all the new calculations together, they constructed annihilation spectra of ISM.  

The fair question is how far the positrons of energy about  one MeV can travel from the Galactic central region to be annihilated. In order to answer to this question one has to compare 
the diffusion (travel) time with the energy loss (thermalization) time with taking into the main energy loss mechanisms.
In general terms, the particle diffusion is related to escape of a magnetized turbulent plasma. 

Jean et al. (2005) present a spectral analyzis of the $e^+e^-$ annihilation emission from the GC region based on the first year of measurements made with the
spectrometer SPI of the INTEGRAL mission. They also analyze the positron  diffusion propagation in the GC region in detail.
The quasilinear diffusion coefficient  
the diffusion coefficient $D_{ql}$ (Melrose 1980) can be expressed as: 
\begin{equation}
D_{ql}=D_{\rm B}\left(\frac{r_{\rm L}}{\lambda_{max}}\right)^{1-\delta}\eta^{-1}
\label{d_ql}
\end{equation} 
with $D_{\rm B}= (1/3) v r_{\rm L}$ the Bohm diffusion coefficient, $\lambda_{max}$ is the maximum scale of the turbulence, $\delta=5/3$ for a Kolmogorov
turbulent spectrum, $\eta=\delta B^2/<B^2>$ is relative perturbation of the magnetic field pressure which is often approximated to 1, 
 $v$ is the positron velocity and $r_{\rm L}$ is the Larmor radius (gyroradius)
\begin{equation}
r_{\rm L}=\frac{m_ec\gamma v}{qB},
\label{r_l}
\end{equation}
 $q$ is the positron (electron) charge,  $B$ is the magnetic field strength, $\gamma=[1-(v/c)^2]^{-1/2}$.
The maximum scale $\lambda_{max}$ was estimated to be $\sim 100$ pc from the measurements of ISM turbulence (Armstrong et al. 1995). 
 
Recently the (mean) magnetic field strength in the GC region was measured by La Rosa et al. (2005), hereafter LaR05, to be $\sim 10$ $\mu$G. 
LaR05 have used the Very Large Array in all four configurations to image the GC region at 74 MHz. 
The resulting image of large scale ($6^o\times2^o$) of nonthermal synchrotron emission  presented in LaR05 has a resolution of $125''$.
At distance  of 8 kpc the angular scales of $6^o\times2^o$ corresponds to region $840~{\rm pc}\times280$ pc and that of $125''$ corresponds to 0.09 pc.
It has be noted that in the past several authors (see Morris \& Serabyn et
al. 1996 and references therein) estimated magnetic fields values of $\sim m$G in the GC region. This is 2 orders of magnitude 
larger than the value ($\sim 10$ $\mu$G) obtained by  LaR05. 

Thus for the mean B of $10$ $\mu$G and for 1 MeV positrons $D_{ql}\sim 2.6\times10^{26}$ cm$^2$s$^{-1}$. 
The low limit time scale for 1 MeV positrons to thermalize in a the $B\sim 10$ $\mu$G, and~~ $\sim 1$ cm $^{-3}$ density region  due 
to ionization, bremsstrahlung, synchrotron and other processes is $\tau \gax 10^5$ years.
Then the distance travelled by diffusion in a time of $\tau_d\sim 10^7$ years (slowing down time plus annihilation time in the standard grain model) is     
\begin{equation}
d_{ql}=\sqrt{D_{ql}\tau_d}=3~{\rm kpc} 
\label{t_d}
\end{equation}
which is much greater than typical half-size  of the hot GC region. The volume filling factor of the hot GC region 0.72 is highest among all phases of the Galactic bulge 
(see more details of these estimates in Jean et al. 2005).
Using the spectral data and positron propagation analysis   Jean et al. (2005) {\it come to conclusion  
that the sources of annihilation emission are diffusively distributed over the Galaxy}.  They also  explain that  the lack of annihilation emission 
from the GC hot gas is due to its low density, which allows positrons to escape this phase.

It is worth noting   Liu, Petrosian \& Melia  (2004) find that an outflow of low energy electrons of order 1 MeV  
are distributed over  a spatially larger scale than that of the BH inner region (see more discussion
of their work in section 6.2).  Namely, the escape time of these electrons is more than a factor of 2 shorter than the time scales for acceleration and losses. 
 
On the hand one can argue that the positrons can be diffusely trapped, thermalized  and annihilated in the region that size $L$ is of 
 order a few parsec and where the magnetic field can be much higher than $\sim 10$ $\mu$G. 
Melia \& Falcke (2001) suggest that out to $\sim2-3$ pc (essentially inside the cavity 
surrounded by the circumnuclear disk), the field is turbulent (matching
the turbulent plasma generated by wind-wind interactions from the
Wolf-Rayet and O/B stars in this region), can be as high as   $\sim0.2-1$ mG.
 
So the problem is that the gyroradius of the electrons and positrons 
produced in gamma-gamma interactions, will be far smaller than the 
region over which the positrons are believed to be annihilating. 
Consequently, the pairs should be radiating profusely 
via synchrotron, bremsstrahlung, and inverse Compton. 

The photon energy density in this region is about $1$ eV cm$^{-3}$. 
To obtain this number we assume that the bolometric luminosity of the source in the Galactic
Center is of order $10^{38}$ erg s$^{-1}$ (see e.g. Narayan et al. 1998).  Thus the derived magnetic strength $B_{bol}$ using equipartition between photons and magnetic field
energies is about $10$ $\mu$G. For  such a low magnetic field we have already demonstrated that the positrons can escape from the GC region.  

In fact, Longair (1994)  derived the minimum magnetic energy requirement for a given luminosity ${\cal L_{\nu}}$ and photon volume V per unit time
(in our case $V=4\pi L^2 c)$. He shows that $B_{min}\propto ({\cal L_\nu}/V)^{2/7}$  which value is higher than that obtained using relation 
${\cal L_{\nu}}/(4\pi L^2 c)=B_{bol}^2/8\pi$.
But the ratio of $B_{min}/B_{bol}$  is not  orders of magnitude. The difference of the indicies of ratio ${\cal L_\nu}/V$ is $2/7$ vs $1/2$.
Furthermore Longair found that the magnetic field strength, $B_{min}$, corresponds to approximate equality of the energies in the particles and magnetic field.
LaR05 confirm this conclusion.  They show 
that the mean magnetic field in the GC region $10$ $\mu$G inferred from their observations is consistent with the particle energy density of  $1.2$ eV
found in cosmic-ray data.   

The  turbulent diffusion time scale $\tau=L^2/D_{ql}$ is not determined
by the magnetic field only.  In fact for $\delta=5/3$ the diffusion coefficient $D_{ql}$ is proportional to $B^{-1/3}$ (see Eq.\ref{d_ql}).
Even if the magnetic field is about $1$ mG the diffusion coefficient $D_{ql}$ decreases by factor 4.6 only, 
namely $D_{ql}\sim 5.5\times 10^{25}$ 
cm$^2$s$^{-1}$. The related difussion time through this region $\tau$ is about $5\times10^4$ years
 which is much less than the thermalization and annihilation time $10^7$ years.


\section{${\bf}\gamma$-Emission as the disk emission in small mass black holes}
One of the possible origins for MeV-emission  is a disk emission from 
mass accretion by SMall Mass Black Holes (SMMBHs). The BH mass 
can be evaluated using the color disk temperature $T_{col}$. Using formula (5)
in Shrader \& Titarchuk (1999), hereafter ShT99, we obtain that
a black hole mass in solar units ($m=M_{bh}/M_{\odot}$)
\begin{equation}
m=\dot m \frac{(1/7)3^4T_h^4}{(T_{max})^4[(7/6)^2r_{in}]^3}
\label{bhmass}
\end{equation}
where  $\dot m=\dot M/\dot M_{\rm Edd}$ is the dimensionless mass accretion rate  in units of $\dot M_{\rm Edd}=L_{\rm Edd}/c^2$,
related to the Eddington luminosty $L_{\rm Edd}$,  
$T_{max}\approx 1.2 T_{col}$  is the maximum temperature in the
disk in keV,
$T_h=T_{col}/T_{eff}$ is a disk color factor and $r_{in}$ is the inner disk
radius calculated in Schwarzchild units. 
Analyzing quite a few BH sources, ShT99 and Borozdin et al. (1999), hereafter
BRT99, established 
that in the soft state the inner disk radius is very close to the last stable
Keplerian orbit, namely $r_{in}\approx3$. 
The disk color factor $T_h\approx 2.6$ was calculated by BRT99 using 
the known contraints on BH mass and the distance to the source for GRO J1655-40.
This value of the disk color factor $T_h\approx 2.6$ obtained for GRO J1655-40 was recently confirmed for other Galatic black hole candidates sources.
Shrader \& Titarchuk (2003) have made use of improved Galactic black hole binary dynamical mass determinations to derive, in effect, 
an empirical calibration of this factor.   

In the soft  state the mass
accretion rate in the disk $\dot m=\dot M/\dot M_{\rm Edd}$ is  in the order
of Eddington, i.e $\dot m\approx 1$. 
With assumption regarding $r_{in}$,  $T_h$ and $\dot m$ we obtain 
\begin{equation}
m=3.7\times 10^{-16}(\dot m/1) \frac{(1/7)3^4(T_h/2.6)^4}
{(T_{max}/1.2\cdot10^4 {\rm keV})^4[(7/6)^2(r_{in}/3]^3}.
\end{equation}

The SMMBHs can be exposed through the accretion, 
if the mass accretion is  at the Eddington rate and higher. 
The Eddington accretion rate by definition is 
\begin{equation}
\dot M_{Edd}=L_{Edd}/c^2=1.4\times10^{17}m~{\rm g~s^{-1}}.
\end{equation}
Thus for   $m=m_{smmbh}=3.7\times10^{-16}$   the Eddington rate is 
about 52 g s$^{-1}$. 
The gravitation attraction of the black hole in the Galactic center
causes  a substantial mass accretion in its proximity.  This
accretion flow passing through SMMBH which are probably
distributed uniformly around the central black hole, 
provides enough material for mass supply of the SMMBH objects. 

Sgr A$^{\ast}$, the massive BH in the Galactic Center was recently found
to be surrounded by a cluster of young, massive stars. Over the past
year, three of these stars named SO-2, SO-16 and SO-19 by Ghez et al
(2004) and Schodel et al (2003) are at $\sim 10^3$ Schwarzschild radii
from the central black hole. It was also found that these stars can be
classified as O stars of the main sequence (Ghez et al. 2003; Eisenhauer
et al. 2003). The winds from these stars have a typical mass loss $
{\dot M}_{w}\sim 10^{-6}M_{\odot}$ yr$^{-1}$ and speed in the range
$v_w=(1-3)\times10^3$ km s$^{-1}$ (Puls et al. 1996; Repolust, Puls \&
Herrero 2004). Loeb (2004) showed that SO-2, SO-16, SO-19 can supply
required mass flow that fuels the emission from Sgr A$^{\ast}$. 

Now we can check if this mass accretion rate $
{\dot M}_{w}\sim 10^{-6}M_{\odot}$ yr$^{-1}\sim 10^{20}$ g s$^{-1}$ can provide
the Eddington mass accretion into SMMBH of $\sim50$ g s$^{-1}$. In order to
do this one should calculate the SMMBH luminosity of
$m_{smmbh}=3.7\times10^{-16}$ for a given color
temperature $T_{col}=10^4$ keV as follows (BRT99)
\begin{equation}
L_{smmbh}=1.4\times10^{34}(\pi^5/15)m^2r_{eff}^2[T_{col}({\rm keV})/T_h]^4~{\rm
ergs~s}^{-1}\approx 2\times10^{21}{\rm ergs~s}^{-1}
 \end{equation}  
where $r_{eff}\sim 15$ is the disk effective radius in Schwarzschild units, and the color factor $T_h\sim 2.6$ (see BRT99).

Then we can evaluate the number of SMMBH $N_{smmbh}$ in the GC region if we assume they are
responsible for $\gamma-$ray emission about  $4\times 10^{38}$ ergs~s$^{-1}$,
namely $N_{smmbh}=4\times 10^{38}/2\times10^{21}=2\times10^{17}$.  One can
conclude that the mass
flow rate $\dot M_{smmbh}N_{smmbh}=10^{19}$g~s$^{-1}$  needed to
fuel SMMBH in the
Galactic Center region within $10^{14}$ cm (or within $\sim 10^{3}$
Schwarzschild radii from the central BH) is much smaller then the mass
flow $10^{20}$ g s$^{-1}$ supplied by the young massive O-stars in this central region. 
Consequently, {\it we show how the small mass black holes could be exposed 
when there is enough  material around them}.

\section{Discussion}
\subsection{The spectrum and  origin of X-ray emission of a BH in the Galactic Center}
 Revnivtsev et al. (2004) claimed that the inferred  X-ray spectrum and luminosity of the emission illuminating the molecular cloud Sgr B2's,  
strongly support the idea that the X-ray emission of Sgr B2 is Compton scattered and reprocessed radiation emitted in the past by the Sgr A$^{\ast}$ source.
We have  calculated the pair creation using  the inferred luminosity and the inferred spectrum that is a power-law photon spectrum of index 1.8 with an exponential cutoff 
at 100 keV. In fact, this is a typical spectrum of BH sources at low/hard state which is presumably a result of Comptonization of the soft (disk) radiation
in the hot electrons of temperature 50 keV in Compton cloud around central BH.

It has already been shown that the spectral shape in the low hard state is independent of the luminosity (e.g. Laurent \& Titarchuk 1999, Titarchuk \& Fiorito
2004). We can propose that   Sgr A* is now in a quiescent state, presumably a result of the gravitational energy release
in the advection dominated flow (see e.g Narayan \& Yi 1994). 

The luminosity  of the X-ray spectrum of the X-ray source IGR J17475-2822 does not change since GRANAT/ART-P and ASCA era, i.e.
during last 10 years. The clear match of ASCA/GIS, GRANAT/ART-P and INTEGRAL/IBIS (Fig. 2, Revnivtsev et al. 2004) strongly supports this claim.  
Thus  the proposed incident spectrum emanating from Sgr A* in the past (300 years ago) should not be vary either 
during the period at least of 10 years. It means that the inferred e+e- rate should be quite stable during the same period.
The constraints on the ``breathtaking variability on a time scale of only several hundred  years'' was proposed by Sunyaev et al. (1993).
They argue that X-ray echo from Sgr B2 should be delayed by 300-400 years relative to the direct signal from Sgr A$^{\ast}$ due to the light travel 
time from Sgr A$^{\ast}$ to Sgr B2. One can put the fair question regarding  such a short recurrence time  for the spectral transition, 
as 300-400 years for a black hole of mass $\sim 3\times10^6$ solar masses.



On the other hand one can suggest  the more 
likely scenario for irradiating Sgr B2 was the interaction between 
the supernova remnant, we now see as Sgr A East and the $50$ km s$^{-1}$ molecular 
cloud behind Sgr A$^{\ast}$, rather than Sgr A$^{\ast}$ itself. The luminosity, time
scale, and spectrum all fit the requirements rather well, so the need
for a high X-ray activity in Sgr A$^{\ast}$ can be  weaker now.
This effect of supenova (SN) ejecta  and its interactions with molecular clouds proposed by Bykov (2002, 2003) is recently discussed in the literature
(see e.g. Muno et al. 2004b,  and Park et al. 2004). One can be right that ``the luminosity, time
scale, and spectrum all fit the requirements rather well''. However Park et al. pointed out  that although the SN ejecta contribution
to the observed neutral Fe line emission appears to be plausible, they noted a caveat, that nonthermal radio emission features have not been
reported in the northeast regions of Sgr A$^{\ast}$.
 
This absence of nonthermal radio features is  admittedly a difficult problem
in terms of the SNR interpretation.  It means that the supernova ejecta model needs the nonthermal distribution of electrons in order to
explain the hard X-ray  energy spectrum of the Sgr B2 molecular cloud (Revnivtsev et al. 2004). 
We want to emphasize once again that the spectrum seen from Sgr B2 is a typical spectrum of the low/hard state
seen in many BH sources. It has already been proven long  ago (see e.g. Sunyaev \& Titarchuk 1980) that this hard spectrum is a result of
thermal Comptonization of soft (disk) photons in the hot Compton cloud surrounding the central part of a BH. 
Thus one does not need   any fine tuning or nonthermal electron distribution to produce such a spectrum,
because in the outskirts of the GC BH there are plenty of molecular clouds, and we should see the effect of the reflection of the BH X-ray spectrum 
from these configurations. Thus  the spectra of molecular clouds are naturally formed as
a result of reflection of BH emission from molecular clouds. 
   
\subsection{The site and the origin of  the annihilation line}
In connection with the previous points, one can raise a question related to expected
rapid variability in the $e+e-$ annihilation line flux for such a compact BH
region.   However the observations over several decades seem to not show any
variability at all.
  
It is correct to say that the variability in the $e+e-$ annihilation line flux would be quite high 
 if this line was formed in  the  compact central region. However, we propose that the pairs are produced in this compact region and only then  do they
   propagate through the Galaxy and   the e+e- annihilation line is formed through the Galaxy.  Thus variability of the e+e- annihilation line flux is related to the
   scale of a few kpc and it is not by chance that  observations over several decades seem to not show any
   variability at all.

We infer that the $\sim$MeV gamma ray region, where the pairs are created due to X-ray illuminations by the central source, is quite compact. 
Its radius is $3.2\times10^{14}$ cm.  
This region within $10 ^{14}$ cm, is consistent with the COMPTEL map (Schonfelder et al. 2000, Strong et al. 1998).
The full sky intensity map,  longtitude and latitude profiles for 1-3, 3-10 and 10-30 MeV show the strong peaks within a few degrees near GC.
However, the spatial resolution of the COMPTEL map is not better than one degree and thus the compact region of $\sim 10^{14}$ cm cannot be resolved 
in COMPTEL map.   It is worth noting that Strong \& Moskalenko (1999) argue that  there is an unresolved point-source population  
in the inner Galaxy and that this source is an important contribution to  the emission around 10 MeV.

Producing $\sim 10^{42}$ s$^{-1}$  (or more, depending on the duty 
cycle) has consequences beyond simply calculating the annihilation 
rate.  The region surrounding Sgr A$^{\ast}$ is magnetized, with magnetic field strength $B$ of order 
10 G inside $10^{14}$ cm . Out to $\sim2-3$ pc  the field is turbulent (matching
the turbulent plasma generated by wind-wind interactions from the
Wolf-Rayet and O/B stars in this region).  Melia \& Falcke (2001) have suggested that the magnetic field strength of this region  should be in the range 
of $\sim0.2-1$ mG.    

In section 4.2 we  review the positron propagation in the Galactic Center region and we find the positrons can  pass through  there.
Consequently the sources of annihilation emission are presumably distributed over the Galaxy.


 Recently Liu, Petrosian \& Melia (2004), hereafter LPM04, formulated and solved 
 the propagation and acceleration problem in the turbulent magnetized plasma 
for a given pair flux $Q$, magnetic field strength $B$, plasma interaction time 
$\tau_p$, density  $n$, the size of the acceleration site  $R$ and particle
 distribution of the background plasma.. They showed  that acceleration of electrons
 by plasma wave turbulence in hot gas  in the inner-most part of the black hole reasonably
 account for Sgr A$^{\ast}$'s millimeter and shorter wavelength emission in the
 quiescent state and for the infrared and X-ray flares. 
 In their model A ($Q=3.2\times10^{42}$ s$^{-1}$, $B\sim 9$ G) 
 they found the power carried away by accelerated 
 electrons, with $\gamma>100$, is about $2\times10^{37}$ ergs s$^{-1}$; which is
 more than enough to power observed radio emission, whose liminosity of $10^{34}-10^{35}$
 ergs s$^{-1}$.   

For the pair flux of order $10^{42}$ s$^{-1}$, 
the ``collateral'' spectrum (and flux)  arising from these particles is very consistent with Sgr A$^{\ast}$'s millimeter and shorter wavelength emission 
in the quiescent state and for the infrared and X-ray flares (LPM04). It is worth noting that the low energy electron flux $E\gax mc^2$ calculated by LPM04, is
 comparable with the injected flux $Q$. It means that a significant fraction of the low enery electrons escapes from this region without any noticeable
 acceleration. In fact, the escape time of low energy electrons is more than a factor of 2 shorter than the time scales for acceleration and losses. The
 exact value of the electron flux depends on the model assumptions (see LPM04); and thus these theoretical uncertainties of the flux estimate 
 (within a factor of a few) are unavoidable. 
 The LPM04's model also suggested that an outflow of high energy electrons (Liu \& Melia 2002)
 are distributed over  a spatially larger scale than that of the BH inner region. Finally the escaping high energy electron gas (cooled through the
 radiation)  could form the annihilation line emission in  the large scale of the Galaxy. 
 The LPM04's injected flux $Q\sim 10^{42}$ s$^{-1}$,  is related
 to the mass accretion rate of order $10^{-5}$ of the Eddington value ($\dot M_{\rm Edd}=L_{\rm Edd}/c^2$). 
  If the pairs are  accelerated to $\gamma>100$ then their energy would be
 enough to power the observed radio emission  whose luminosity is about $10^{34}-10^{35}$ ergs s$^{-1}$.
 
 It is worth noting that  within our model, the same flux of  pairs ($Q\sim 10^{42}$ s$^{-1}$)  is generated in the SMMBH site as a result of X-ray illumination emitted 
 by surrounding material. If at the present time  the X-ray flux is of order $10^{36}$ erg s$^{-1}$, then the pair flux  would be $10^{39}$ s$^{-1}$.
 
   
 \subsection{The dark matter problem and its relation to  observed annihilation line flux}
Dark matter is a long standing puzzle starting from  Le Verrier
with his discovery of Neptune in 1846 as a perturbation of Uranus
(Le Verrier 1846).
Our Galaxy is surrounded by an extended halo of unseen material.
So far undetected, that dark matter induces a flat rotation curve in the
Galactic plane  (van Albada \& Sancisi 1986).
Its presence has been noticed since many decades (see Oort  1932).
However, its nature is still unknown. The dark matter problem consists simply, in the
existence of invisible mass showing its presence by gravitational effects.  Now, it is  widely  accepted 
that dark matter exists. In fact, there is  evidence for dark matter on scales 
from galaxy to the cluster of galaxies and to the whole Universe itself.  
 Zwicky  (1937) measured the dispersion velocity in the Coma
cluster and found that the dynamical mass was hundred times more than the
luminosity mass.

Moreover, the dark matter seems essential: the growth of
structure in the Universe by hierachical
merging of the dark matter halo is a master piece of moderm cosmology.
 Recently, WMAP has established
the presence of non baryonic dark matter with the density reported in 
terms of critical density of order $\Omega_{m}=\rho_{m}/\rho_{c}\simeq  0.3$. 
This is now the standard value of the modern cosmology for dark 
matter density.

On the galactic scale, one can interprete the rotation curve measurements by 
the presence of the dark matter halo and then determine  the mass density profile of this
distribution. The average density is $\rho_{m} \simeq 0.3 \; {\rm GeV}~ {\rm cm}^{-3}$. 
The core of this halo seems to be extremely packed. In the Galactic Center, numerical simulations of
structure formation in the non-linear regime have shown the presence
of a singular power law cusp $\rho(r) = \rho_0 (r/r_0)^{-\gamma}$,  where $0 <  \gamma\lax 2$ and  $\rho_0$ is 
the dark matter density at radius $r_0 \simeq  2 $ pc  (Gnedin \& Primack 2004).
The central density $ \rho_{0}$ is not uniquely determined and its
value depends on models and it varies from 60 to 600 $M_{\sun}$ pc$^{-3}$. 
On the other hand up to now there is no identification for it despite the
large number of candidates proposed from masses in the range between 
$10^{-5}$ eV to $10^{12} \; M_{\sun}$. 

While there is no a priori reason to expect that massive compact halo 
objects (MACHOs) in the mass range $10^{-15} M_{\odot } \lax M <
10^{-7} \; M_{\odot }$ comprise a significant fraction of the density of
the Universe, neither is there any definitive argument ruling them out.
Nemiroff \& Gould (1995) and  Marani et al. (1999)  discuss the
possibility of identification of SMMBHs by gravitational
lensing. 
The basic idea is to study the deflection of light by measuring the
amplification due to gravitational bending of compact objects.
For black hole mass in the range $10^{-16} \; M_{\odot }$ to
$10^{-13} \; M_{\odot }$, the expected image separation is of order of
$10^{-15}$ arcseconds and the average time separation between the
images is order of $10^{-24}$ s.  They applied this delay as an
interference process using gamma ray bursters.
On the other hand, the absolute value of the light deflection by small black holes is extremely
small $(10^{-12}~{\rm cm})/R$, where R is the characteristic  distance in the Galaxy 
and therefore  not observable.   It is  easy to show 
that the optical depth for the photons scattering at small angles
by SMMBH is  small,  ($\sim 10^{-6}$) and consequently the light multiple scattering  by the small black holes   
effect  is also ruled out. 

The supersymmetry theory  as an extension of the Standard Model, predicts that one stable particle
should be around today as a cosmological relic.
 Among these new particles, the neutralinos, could be the dark matter. 
If the dark matter halo of our Galaxy is made up of such 
particles, they will annihilate each other into fermion-antifermions
pairs, and this will produce stable particles which could be detected
as anti-protons, electrons and photons. Another way is to look for
direct detection via laboratory experiments. Such
new particles are scrutinized by direct  and indirect searches since more 
than ten years.
Our Galaxy is indeed a subject of intense activities because it could be a
direct search for dark matter in studying astrophysical signatures. 

Among all the signatures, our Galaxy presents three strong signatures 
in X-ray, $\gamma-$ray backgounds and a typical $e^+~e^-$ annihilation source in the
Galactic Center.  Among the other possibilities, the micro-lensing searches have been
developed and used to probe the dark matter halo of our galaxy (see above).
 The possibility that the halo could be made of primordial black holes
 is not ruled out by observations. 
 Recently, Seto and Cooray (2004) discussed also the future
 possibility of searching dark matter candidates using the space-based 
 gravitationnal wave detectors. 
 They used this idea to test
 the possibility that the halo could be made of primordial black holes.
According to Hawking predictions such primordial black holes will not
suffer evaporation if their mass is greater than $10^{15}$ g.

While the consequences of evaporating  black holes has
been widely analyzed in terms of particles production, the case for
stable SMMBHs s has not been considered so far. In
literature, the problem is analyzed generically and the mass range used are often very
large from $10^{17}$ g  to $10^{20}$  g. 

\section{Conclusions} 
In this Paper, we show that there is a particular 
  black hole mass which is very interesting for the dark matter problem.
This mass of order of  $10^{17}$ g, corresponds to a typical Schwarzschild
radius  proton size order.  We explore this possibility 
in the light of existing galactic diffuse X and $\gamma$ backgrounds and 
with the strong source of annihilation lines at 511 keV in the
Galactic Center. 
Our approach is to find a natural explanation of such astrophysical
backgrounds  within an astrophysical dark matter scenario. \ 

We also offer a model for the origin of 
the annihilation line appearance in the our Galaxy 
which has been detected in several balloon and space experiments.
We combine X-rays and $\gamma$-observations of the Galactic Center region to
show that pair production can be a  result of interactions of X-rays
and gamma photons emitted in this region.
Pairs production strongly depends on the compactness of the
X-rays and $\gamma$ emission region which is determined by
X-rays and  $\gamma$ luminosity and size of the region.

For velocity of the interstellar medium of much less than $10^8$ cm s$^{-1}$ and the value of density
column a few times $10^{22}$, the optical depth for 511 keV line is
more than one.
We found  for the observed X-rays and $\gamma$ luminosities and for the observed
annihilation lines flux, the size of the X-ray and $\gamma$ emission
area is    $\sim 3\times10^{14}$ cm. This size is consistent with 
the size of the area where the OB stars  supply sufficient amount of fuel
required for the X-rays and $\gamma$ emissions from the Galactic Center
region.

If we compute the total black hole mass inside this region of radius
R=$10^{-4}$ pc, we obtain that this mass is about two solar masses. 
We can then naturally ask if SMMBHs  can be a candidate
for dark matter as already proposed in literature ( e.g. Nemiroff \& Gould 1995,
Marani et al. 1999).  
It is interesting to compare this number to the dark matter mass computed
using the singular power law cusp $\rho(r) = \rho_0 (r/r_0)^{-\gamma}$
often used in the dark matter model for the Galactic Center (see Moore et al. 1998 and Gnedin \& Primack 2004). 
We found that with a set of parameters
($\rho_0 =50 \; M_{\odot }$ pc$^{-3}$, $r_0 =2 \; $ pc , $\gamma = 2 $), 
we can obtain a mass of $0.5 \; M_{\odot}$ inside the same radius R,
which is compatible to our number within a factor of a few.

Koyama (1996) and Muno et al. (2004a), Revnivtsev et al. (2004) noted that the Galactic Center exhibited intermittent activities with time-averaged 
energy generation rate $\sim 10^{41-42}$ ergs s$^{-1}$ comparable to Seyfert nuclei. Muno et al. (2004a) 
are left to conclude either that there is a significant
shortcoming in understanding of mechanisms that heat  the interstellar medium in
Galactic Center region or that of a population of faint
($<10^{31}$  ergs s$^{-1}$) hard X-ray sources.  Muno et al. (2004b) using a deep Chandra observations
study to detail the origin of
the point like sources near the Galactic Center. Possibly a single
population of sources may dominate in this cluster. There are
certainly many classes of objects present in this set of faint
sources. We may suggest   that some part of X-ray and $\gamma-$radiation of the Galactic Center, can  originate in accreting flows
into SMMBHs (that are really faint sources with their bolometric luminosity $<10^{22}$  ergs s$^{-1}$)  
in which outer parts of   flow emit X-rays and the innermost part  of the flow emit
$\gamma-$rays.

Finally we can conclude, that the only way to identify small mass black holes (SMMHHs) is by accretion
mechanism in presence of a surrounding dense material. In the absence of
this matter these SMMBHs are invisible and can be treated as
 dark matter, or invisible matter, i.e. without any observational appearance in the 
radiation.
 
We appreciate productive  discussions with Gerald Shore and Robert Duffin.
For this work, L.T.  was partly  supported by French  Minist\`ere de l'Education Nationale et de la
Recherche with grant SSHN2004-413077D. We thank the referee for his/her profound questions and we 
acknowledge the referee's contribution in the discussion section of this paper.





\newpage
\begin{figure}[c]
\includegraphics[width=5.4in,height=9.in,angle=0]{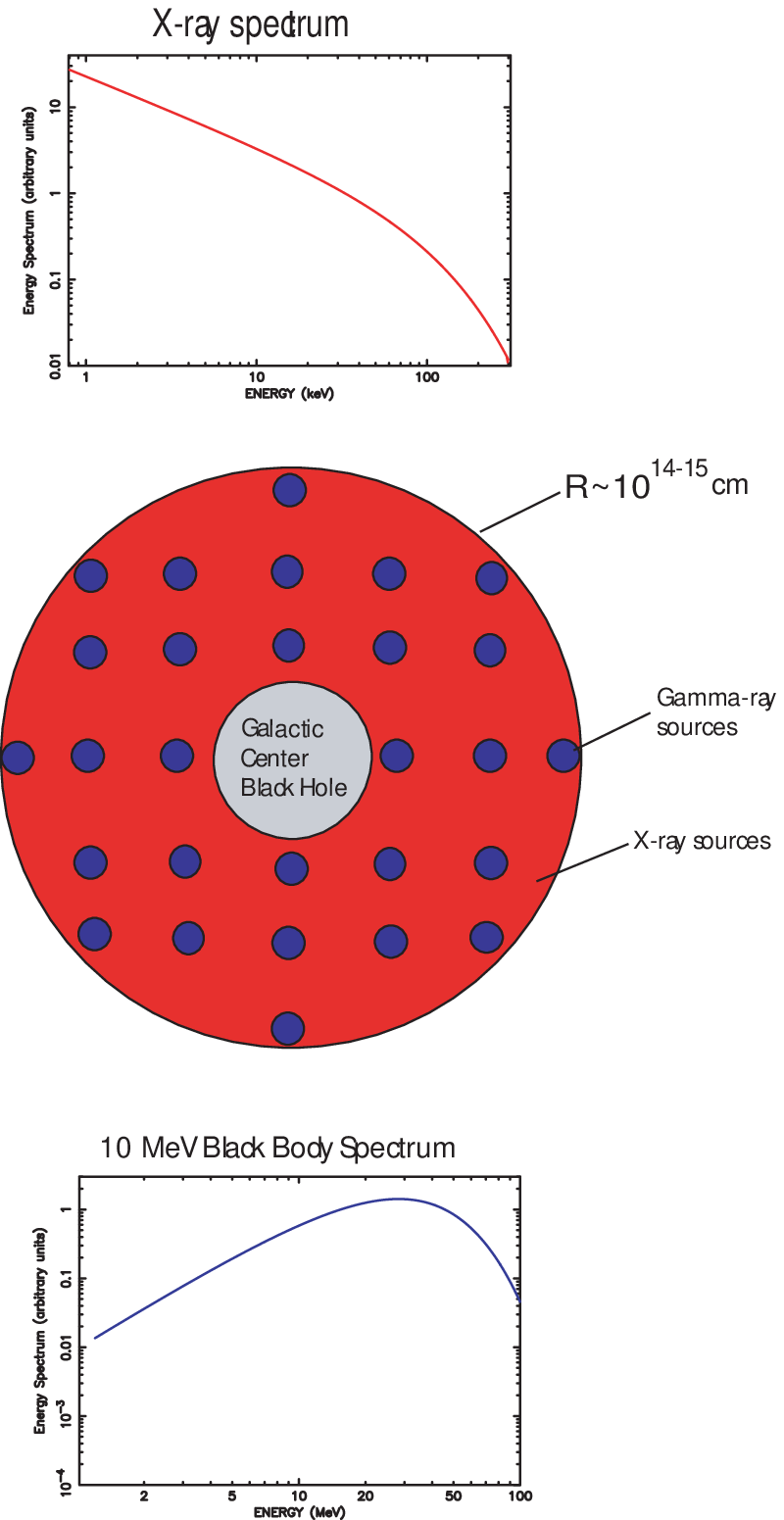}
\caption{ Cartoon picture of compact emission area near the Galactic
Center. Sources of $\gamma$-ray region in blue,
and sources of X-ray radiation in red. In upper and lower panels
we show    X and $\gamma$ rays spectra respectively.   }
\label{picture}
\end{figure}


\begin{thebibliography}{}


\bibitem[Aharonian et al. (2004)]{ahar} Aharonian, F., et al. 2004, A\&A, 495, L13
\bibitem[Armstrong et al. (1995)]{arm} Armstrong, J.W., Rickett, B.J. \& Spangler, S.R. 1995, \apj, 443, 209
\bibitem[Bananoff et al. (1997)]{bagan} Baganoff, F.K., 2003, \apj, 591, 891
\bibitem[Boehm et al. (2004)]{boeh} Boehm, C, Fayet, P., \& Silk, J. 2004, Phys. Rev, D69, 101302
\bibitem[Bykov (2003)]{byk03} Bykov, A.M. 2003, A\&A, 410, L5
\bibitem[Bykov (2002)]{byk02} Bykov, A.M. 2002, A\&A, 390, 327
\bibitem[Borozdin et al. (1999)]{bor}
Borozdin, K., Revnivtsev, M., Trudolyubov, S., Shrader, C, \&  
Titarchuk, L.  1999, ApJ, 517, 367 (BRT99)
\bibitem[Chakrabarti \& Titarchuk (1995)]{ct95}
Chakrabarti S.K. \& Titarchuk, L. G. 1995, ApJ,  455, 623 
\bibitem[Chapuis et al. (1991)]{chap} Chapuis, C.G.L. et al. 1991, Gamma-Ray Line Astrophysics, ed. P. Durouchoux \& N. Prantos
(New York: AIP), 54
\bibitem[Cheng et al. (1997)]{cheng97} Cheng, L.X., 1997, \apj, 481, L43  
\bibitem[Chez et al. (2004)]{chez04} Chez, A.M., et al.  2004, \apj, 601, L159
\bibitem[Chez et al. (2003)]{chez03} Chez, A.M., et al.  2003, \apj, 586, L127
\bibitem[Churazov et al. (2005)]{chur} Churazov, E., Sunyaev, R., Sazonov, S. \&  Varshalovich, D. 2005, MNRAS, 357, 1377 
\bibitem[Dermer \&  Skibo  (1997)]{Dermer97} Dermer,C.D. \&  Skibo, J.G.  1997, \apj,    487, 57
\bibitem[Dirac  (1930)]{dirac} Dirac, P.A.M.   1930, Proc. Camb, Phil. Soc.   26, 361
\bibitem[Eisenhauer et al. 2003]{eisen} Eisenhauer, F., et al. 2003, \apj,   597, L121
\bibitem[Gehrels et al. (1991)]{gehr} Gehrels, N., et al. 1991, \apj, 375, L13
\bibitem[Gnedin \& Primack (2004)] {Gnedin04} Gnedin, O.Y., \& Primack, J.R. 2004,  {\sl Phys. Rev. Lett. }, 61, 302 
\bibitem[Guessoum et al. (2005)]{gue} Guessoum, N., Jean, P \& Gillard, W. 2005, A\&A, 436,171
\bibitem[Heitler (1954)]{heitler} Heitler, W. 1954,  The Quantum Theory of Radiation, 
Oxford University Press,  Oxford
\bibitem[Hunter et al. (1997)]{huntb} Hunter, S.D., Bertsch, J.R., Catelli, J.R. 1997, in AIP Conf Proc. 410, 4th Compton Symposium, New York, 1193
\bibitem[Jean et al. (2005)]{Jean05} Jean, P.   2005, A\&A  in press (astro-ph/0509298)
\bibitem[Kawai et al. (1988)]{kaw} Kawai, N., et al.  1988, \apj, 330, 130 
\bibitem[Koyama et al. (1996)]{koy96} Koyama, K.,  et al. 1996, PASJ, 48, 249
\bibitem[Koyama et al. (1989)]{koy89} Koyama, K.,  et al. 1989, Nature, 339, 603
\bibitem[Laurent \& Titarchuk (1999)]{lt}Laurent, P. \& Titarchuk, L. 1999, \apj, 511, 289 
\bibitem[Le Verrier (1846)]{LeVerrier} Le Verrier, 1846, {\sl
Astronomische Nachrichten}, 25, 65
\bibitem[Leventhal et al. (1993)]{leven} Leventhal, M. et al. 1993, \apj, 405, L25
\bibitem[Leventhal et al. (1978)]{lms} Leventhal, M., MacCallum, C.J., \& Stang, P.D.  1978, \apj, 225, L11  
\bibitem[Liu et al. (2004)]{lpm} Liu, S.,  Petrosian, V., \& Melia, F. 2004,  \apj, 611, L101  
\bibitem[Liu \& Melia (2002)]{lm} Liu, S.,  \& Melia, F. 2002,  \apj, 573, L23 
\bibitem[Loeb   (2004]{Loeb2004} Loeb, A. 2004, \mnras, 350, 725
\bibitem[Longair  (1994)]{long} Longair, M.S. 1994, High Energy Astrophysics, 
Cambridge: University Press
 \bibitem[Marani et al. (1999)]{maran} Marani, G.F., Nemiroff, R.J., Norris, K. \& Bonnell, J.T. 1999, \apj, 512, L13
\bibitem[Markevitch et al. (1993)]{mark} Markevitch, M., Sunyaev, R.A.,  \& Pavlinsky, M. 1993,  Nature 364, 40 
\bibitem[Melia  (2001)]{melia} Melia, F. \&  Falcke, H. 2001, ARAA, 39, 309
\bibitem[Melrose  (1980)]{melrose} Melrose, D.B. 1980, Plasma Astrophysics. Nonthermal Processes in Duffuse Magnetized Plasmas, 
New York: Gordon and Breach
\bibitem[Moore et al. (1998)]{moor} Moore, B.,   et al. 1998, \apj, 499, L5
\bibitem[Morris \& Serabyn (1996)]{mors} Morris, M., \& Serabyn, E.   1996, ARA\&A, 34, 645
\bibitem[Muno et al. (2005)]{muno05} Muno, M., et al., 2005, \apj, 622, L113
\bibitem[Muno et al. (2004a)]{munoa} Muno, M., Baganoff, F.K, Bautz, M. W.  et al.
2004a, \apj, 613, 326
\bibitem[Muno et al. (2004b)]{munob} Muno, M., Arabadjis, F. K, Baganoff,
F.K,   et al. 2004b, \apj, 613, 1179
\bibitem[Murphy et al. (1987)]{murphy} Murphy, R.J., Dermer, C.D., \& Ramaty, R. 1987, ApJS, 63, 721
\bibitem[Oort  (1932)]{OORT} Oort J.~H., 1932, {\sl Bull. Astron. Inst. Netherlands} 6, 349
\bibitem[Narayan \& Yi  (1994)]{ny98} Narayan, R., et al.   1998, \apj, 492, 554
\bibitem[Narayan \& Yi  (1994)]{NY94} Narayan, R. \& Yi, I.  1994, \apj, 428, L13
\bibitem[Nemiroff \& Gould  (1995)]{NG} Nemiroff, R.J. \& Gould, A.  1995, \apj, 452, L111
\bibitem[Niel et al. (1990)]{niel} Niel, M. et al. 1990, \apj, 356, L21
\bibitem[Park, et al.  (2004)]{park} Park, S. et al. 2004, \apj, 603, 548
\bibitem[Parker (1979)]{parker} Parker, E. N. 1979,  Cosmical Magnetic Fields, 
Clarendon Press,  Oxford
\bibitem[Pavlinsky,  et al. (1993)]{pavl} Pavlinsky, M., Grebenev, S.A., \& Sunyaev, R.A.,  \&  1993, \apj, 425, 110 
\bibitem[Preddehl, \& Trumper (1994)]{pred} Preddehl, P., \& Trumper,J.  1994, A\&A, 290, L29
\bibitem[Puls et al. (1996)]{puls} Puls et al. 1996,  A\&A, 305, 171
\bibitem[Purcell  et al. (1997)]{Purcell97} Purcell, W.R.,  et al. 1997,  \apj,  491, 725 
\bibitem[Purcell  et al. (1993)]{Purcell93} Purcell, W.R.,  et al. 1993,  \apj,  413, L85
\bibitem[Ramaty et al. (1994)]{ram94} Ramaty, R., Skibo, J.G.  \& Lingenfelter 1994, ApJS, 92, 393  
\bibitem[Repolust, Puls \& Herrero (2004)]{rep04} Repolust, Puls \& Herrero 2004, A\&A, 415, 349  
\bibitem[Revnivtsev et al. (2004)]{rev04} Revnivtsev, M.G., et al. 2004, A\&A, 425, L49 
\bibitem[Schodel et al. (2003)]{schod03} Schodel, R.,   et al. 2003,  \apj, 596, 1015
\bibitem[Schonfelder et al. (2000)]{schon00} Schonfelder, V.,   et al. 2000,  A\&ASS, 143, 145
\bibitem[Seto \&  Cooray (2004)]{Seto} Seto, N. \&  A. Cooray, A. 2004,  {\sl Phys. Rev.} {\bf D}, 70,  63512
\bibitem[Share et al. (1990)]{share90} Share, G.H., et al. 1990, \apj, 358, L45
\bibitem[Shrader \& Titarchuk (2003)]{shT03} Shrader, C., \& Titarchuk, L.G. 2003,  ApJ, 598, 168
\bibitem[Shrader \& Titarchuk (1999)]{shT99} Shrader, C., \& Titarchuk, L.G. 1999,  ApJ, 521, L121 (ShT99) 
\bibitem[Skinner et al. (1987)]{skin} Skinner, G.K., et al. 1987, Nature, 330, 544
\bibitem[Smith (2002)]{smith} Smith, D.M., Heindl, W.A. \& Swank, J.H. 2002,  \apj, 569, 362
\bibitem[Strong \& et al. (1998)]{Str98} Strong, A.M., Bloemen, H., Diehl, R., Hermsen, W., \& Schonfelder, V. 1998, Proc. 3rd. INTEGRAL Workshop (Eds. G. Palumbo, A.Bazano,  
\& C. Winkler) Astrophys. Lett. Comm. 39, 209
\bibitem[Strong \& Moskalenko (1999)]{SM99} Strong, A.M. \& Moskalenko, I.V. 1999, 
Proc. 5th Compton Symposium (Eds.  M. L. McConnell and J. M. Ryan)  AIP Conference Proceedings,  510, 291 
\bibitem[Sunyaev et al. (1993)]{sun} Sunyaev, R.A., Markevitch, M., \& Pavlinsky, M. 1993,  \apj, 407, 606 
\bibitem[Sunyaev \& Titarchuk (1980)]{ST80} Sunyaev, R.A. \& Titarchuk, L.G.  1980,  A\&A, 86, 121 
\bibitem[Takeshima et al. (1991)]{tak} Takeshima, T.  et al. 1991, in proc. of 28th Yamada Conf., Frontiers of X-ray Astronomy, ed. Y.
Tanaka, K. Koyama (Universal Academy Press, Tokyo), 421
\bibitem[Teegarden et al. (1996)]{tee} Teegarden, B.J., et al. 1996, \apj, 463, L75
\bibitem[Tanaka et al. (1995)]{tan} Tanaka, Y., Inoue, H., \& Holt, S.S.  1995, PASJ, 46, L37
\bibitem[Titarchuk \& Fiorito (2004)]{tf04} Titarchuk, L. \& Fiorito, R. 2004,  \apj, 612,  988 
\bibitem[van Albada\& Sancisi (1986)]{ALBADA_SANCISI} van Albada T.~S., \& Sancisi R., 1986,
{\sl Phil. Trans. R. Soc. Lond. A}, 320, 447
\bibitem[Watson et al. (1981)]{watson} Watson, M.G., Willingate, R., Grindlay, J.E., Hertz, P. 1981, \apj, 250, 142
\bibitem[Wood et al. (2001)]{wood} Wood, K.S., et al. 2001, \apj, 563, 246
\bibitem[Yamauchi et al. (1990)]{yam}  Yamauchi, S.  et al. 1990, \apj, 365, 532
\bibitem[Zwicky (1933]{Zwicky} Zwicky, F. 1937, \apj, 86 217






























   

  










  
		


  

		

   








 




































  








  









































\end{thebibliography}
\end{document}